\documentclass[preprint]{emulateapj-rtx4}

\usepackage{rotating}

\usepackage{graphicx}
\usepackage{epstopdf}
\usepackage{natbib}
\usepackage[colorlinks,urlcolor=blue,citecolor=blue,linkcolor=blue]{hyperref} 
\usepackage{xspace}
\usepackage{color}
\usepackage{amsmath}

\newcommand{\kms}    {~km~s$^{-1}$\xspace}
\newcommand{\mjy}    {~mJy~beam$^{-1}$\xspace}

\newcommand{\msun}  {~$M_{\sun}$\xspace}
\newcommand{\lsun}    {~$L_{\sun}$\xspace}

\begin{document}

\title{1.3 MM POLARIZED EMISSION IN THE CIRCUMSTELLAR DISK OF A MASSIVE PROTOSTAR}
\author{M. Fern\'andez-L\'opez\altaffilmark{1}}
\author{I.~W. Stephens\altaffilmark{2,3}}
\author{J.~M. Girart\altaffilmark{4,3}}
\author{L. Looney\altaffilmark{5}}
\author{S. Curiel\altaffilmark{6}}
\author{D. Segura-Cox\altaffilmark{5}}
\author{C. Eswaraiah\altaffilmark{7}}
\author{S.-P. Lai \altaffilmark{7}}

\altaffiltext{1}{Instituto Argentino de Radioastronom{\'{\i}}a (CONICET), CCT La Plata, 1894, Villa Elisa, Argentina; manferna@gmail.edu}
\altaffiltext{2}{Institute for Astrophysical Research, Boston University, Boston, MA 02215, USA}
\altaffiltext{3}{Harvard-Smithsonian Center for Astrophysics, 60 Garden Street, Cambridge, MA 02138, USA}
\altaffiltext{4}{Institut de Ci\`encies de l'Espai, (CSIC-IEEC), Campus UAB, Carrer de Can Magrans S/N, E-08193 Cerdanyola del Valls, Catalonia, Spain}
\altaffiltext{5}{Astronomy Department, University of Illinois, 1002 West Green Street, Urbana, IL 61801, USA}
\altaffiltext{6}{Instituto de Astronom\'\i a, Universidad Nacional Aut\'onoma de Mexico (UNAM), Apartado Postal 70-264, 04510 Mexico, DF, Mexico}
\altaffiltext{7}{Institute of Astronomy \& Department of Physics, National Tsing Hua University, No. 101, Section 2, Kuang-Fu Road, Hsinchu 30013, Taiwan}

\begin{abstract}
We present the first resolved observations of the 1.3~mm polarized emission from 
the disk-like structure surrounding the high-mass protostar Cepheus~A~HW2. These CARMA
data partially resolve the dust polarization, suggesting an uniform morphology 
of polarization vectors with an average 
position angle of $57\degr\pm6\degr$ and an average polarization fraction of 
$2.0\%\pm0.4\%$. The distribution of the polarization vectors can be attributed to 
(1) the direct emission of magnetically aligned grains of dust by a uniform magnetic 
field, or (2) the pattern produced by the scattering of an inclined disk. We show that 
both models can explain the observations, and perhaps a combination of the two 
mechanisms produce the polarized emission. A third model including a toroidal 
magnetic field does not match the observations.
Assuming scattering is the polarization mechanism, these observations suggest that 
during the first few $10^4$ years of high-mass star formation, grain sizes can grow from
$1~\mu$m to several $10$s~$\mu$m. 
\end{abstract}


\section{Introduction}
Recently, polarization has been resolved in a few 
circumstellar disks around low-mass protostars via millimeter and centimeter
radio-interferometric observations. Specifically, these resolved polarization detections 
were found for two Class 0 sources 
\citep[IRAS~16293-2422 and L1527, ][]{2014Rao,2015SeguraCox} and a Class I/II source 
\citep[HL Tauri][]{2014Stephens}. A fourth detection has been reported 
toward the candidate disk of the Class 0 protostar NGC1333 IRAS4A1 \citep{2015Cox}.
However, these observations have a small number of independent polarization
measurements difficulting to discern the nature of the polarized emission.

Usually, the polarized (sub)millimeter continuum emission from 
protostellar disks has been interpreted as a consequence of the presence 
of a magnetic field (although see discussions about other possibilities in \citealt{1952Gold1,1952Gold2,2016Andersson}). 
The short axis of dust grains preferentially aligns with the magnetic field 
causing dust polarization to be perpendicular with the magnetic field
\citep[][and references therein]{2007Lazarian,2016Andersson}. This polarization
mechanism is expected to dominate dust emission at scales larger than the disk,
and is frequently used by interferometers to investigate the role of magnetic fields at
the envelope scale \citep[e.g.,][]{1998Rao,1999Girart,2006Girart,2009Girart,2013Stephens}. 
However, this mechanism is not the only way of producing (sub)millimeter polarized 
emission. Dust grain growth is expected to happen in protostellar disks \citep{2000Beckwith,2014Testi}. 
Scattering by large enough dust grains ($\gtrsim 10~\mu$m) may also cause the observed 
polarized morphologies in the disks, as recently 
shown by \cite{2015Kataoka} and \cite{2016Yang1}. 

The model of \cite{2015Kataoka} proposed that the polarized (sub)millimeter emission 
is produced by scattering of anisotropic radiation in dust grains with 
sizes of $\sim100$s~$\mu$m. \cite{2016Yang1} included the scattering idea to 
dust grains in inclined disks. This study was expanded upon 
in \cite{2016Yang2}, which used  a semi-analytic model to combine 
polarization originating from both scattering and direct emission from grains aligned 
with a toroidal field. The polarization pattern strongly depends on the disk inclination 
\citep[see Figure 3 in][]{2016Yang2}. For a $45\degr$ inclined disk, on one hand, 
the polarized emission due to scattering is concentrated along the disk major axis and the
polarization vectors\footnote{These so-called \emph{vectors} are not actually vectors 
but rather segments showing the orientation of the polarization plane.} 
are generally aligned with the minor axis. On the other hand, for disks 
with such inclination, the polarized direct emission is radial throughout the disk 
(i.e., the polarization vectors are perpendicular to the projected toroidal magnetic field). 
Despite the apparent differences between scattering and polarized dust grain emission, 
the current resolved polarization observations in disks are unable to distinguish clearly 
between them. 

At a distance of 0.7~kpc \citep{2009Moscadelli,2011Dzib}, Cepheus~A~HW2 is one of 
the nearest high-mass YSOs. Its luminosity is about $1.3\times10^4$\lsun \citep{1981Evans}, 
which is that of a B0.5 main sequence star \citep[][although note that because this is not
a main-sequence star some of the luminosity is due to accretion]{1994Rodriguez}. 
\cite{2005Patel} found a rotating molecular and dusty 1-8\msun disk-like structure, 
with a radius of 330~AU \citep{2005Patel,2007Brogan,2007Torrelles,2009JimenezSerra}. 
The dust continuum emission of this structure is elongated 
along the direction perpendicular to a thermal radio jet \citep{1994Rodriguez,1995Hughes,2002Curiel}
which has a $45\degr$ position angle (P.A., measured counterclockwise from north) and an 
ejection velocity of 500\kms \citep{2006Curiel}. 
The radio jet is associated with part of a strong molecular outflow detected in different 
molecular tracers (H$_2$, CO and HCO$^+$). The outflow is precessing due to a 
multiple protostellar system harbored in the disk-like structure of HW2
\citep{2006Curiel,2009Cunningham,2013Zapata}.

Submillimeter and millimeter observations have detected two molecular clumps 
inside the HW2 gaseous disk \citep{2007JimenezSerra,2007Brogan,2007Comito,2007Torrelles} 
which may be interpreted as two separate hot cores \citep{2007Brogan,2008Comito,2009JimenezSerra},
or as gas heated by HW2 \citep[i.e., the central protostar,][]{2007Torrelles}. 

Other radio and millimeter sources in the Cepheus A region are probably harboring
low/intermediate mass protostars. In particular, the protostellar source HW3c is detected 
at 0.8~mm and 1.3~mm at $3\farcs5$ south of HW2 \citep{2007Brogan,2013Zapata}. 
HW3c is probably responsible for a prominent outflow running southwest of its position.

Polarimetric observations with the JCMT\footnote{James 
Clerk Maxwell Telescope} at 850~$\mu$m  ($14\arcsec$ angular resolution) 
show a centrosymmetric pattern of polarization vectors around the HW2/HW3c region 
(e.g., \citealt{2009Matthews,2003Chrysostomou,2007Curran}; also see \citealt{1999Glenn}). 
This pattern is broken 
along a $20\arcsec$ depolarization band stretching across HW2 with 
a P.A.$\sim135\degr$ (i.e., a lane about 10 times the size of the disk-like structure). 
The plane of the sky magnetic field strength 
estimated from these observations is 6~mG \citep{2007Curran}. 

\cite{2006Vlemmings} measured magnetic field strengths 
(through H$_2$O Zeeman-splitting observations at 22.2~GHz) 
in the HW2 area ranging from 30 to 600~mG. These fields are strong
enough to control the outflow dynamics \citep{2007Curran}. Polarimetric observations 
of OH masers at 1665~MHz have a linear vector oriented at 
P.A.$\sim143\degr$ \citep{2005Bartkiewicz}. 
The resolution of the H$_2$O Zeeman-splitting observations and the JCMT polarimetric 
observations are drastically different ($0\farcs0005$ versus $14\arcsec$, respectively), which
likely explains the very different field strengths.
Finally, NIR K-band polarization observations at arcsecond 
resolution \citep{2004Jones} report a centrosymmetric pattern 
in the arcminute scale, which may be interpreted as scattering due to the illumination by the
central star(s). 
This pattern is centered at HW2, and is again disrupted in a southeast-northwest 
elongated lane crossing HW2. In this elongated lane,
the polarization vectors show a roughly uniform direction, approximately aligned with the 
orientation of the lane itself.

In this contribution we present CARMA polarimetric millimeter observations partially resolving 
the circumstellar disk-like structure of Cepheus~A~HW2, which possibly harbors a cluster 
of high-mass protostars. These type of disk structures are generally large reservoirs 
of dense dust and gas, but their polarized emission has yet to be explored in detail.
The paper is organized as follows. 
In Section 2, we describe our CARMA observations. In Section 3, we present the main results and
the spatial distributions from the four Stokes parameters. In Section 4, we briefly discuss 
our findings. Finally, in Section 5, we summarize the main results.

\section{Observations}
The 1.3~mm CARMA observations toward Cepheus~A~HW2 were taken 
on three days: 2014 
March 20, 21 and 2014 April 10. CARMA was in C configuration, 
with baselines ranging between 25~m and 310~m. The phase center 
of the telescope was R.A.(J2000.0)=$20^h 56^m 17\fs98$ and 
DEC(J2000.0)=$62\degr 01\arcmin 49\farcs50$. For these observations, 
the receiver was tuned in Full Stokes mode, with four 500~MHz wide 
bands in each sideband, and a Local Oscilator frequency of 
234.0112~GHz. As gain calibrators we used BLLAC and the quasar 
J0102+584. The bandpass calibrators were 3C454.3 and BLLAC. 
Leakages were solved using BLLAC and J0102+584. The leakage terms for 
each antenna were consistent from track to track, and the accuracy of their 
calibrations is within 0.1\% \citep{2014Hull,2015Hull}. The absolute flux scale was 
determined from observations of MWC349 and BLLAC. We derive their 
fluxes from the CARMA dedicated quasar monitoring. 
A 15\% uncertainty is estimated for the CARMA 
absolute flux calibration of the 230~GHz observations. The three tracks were 
calibrated using the Miriad package \citep{1995Sault}. 
Finally, we combined the data from the three tracks and produced images of the 
four Stokes products ($I$, $Q$, $U$ and $V$). We also create a linear polarization 
intensity image ($p=\sqrt{Q^2+U^2}$) and a polarization position angle 
image ($\chi=0.5~\arctan{U/Q}$, with the arctan calculating the angle in the
appropriate quadrant based on the signs of $Q$ and $U$), constraining 
the polarization detection to those locations in which the linearly polarized 
emission has a signal-to-noise ratio (SNR) greater than 3 and the total 
intensity (Stokes $I$ map) has SNR$>5$. The synthesized beam of the naturally 
weighted images is $0\farcs9\times0\farcs8$ (P.A.$=-48\degr$), about 630~AU at 700~pc. 
The rms noise of the Stokes $I$ map, which is limited by dynamic range, is 10.4\mjy and for 
Stokes $Q$ and $U$ maps, which is limited by thermal-noise, is 1.2\mjy.

\section{Results}
\begin{figure*}[!h]
\epsscale{1}
\plotone{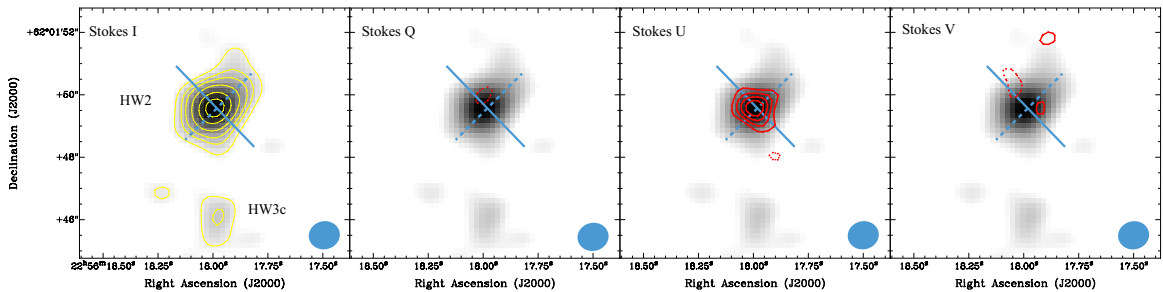}
\caption{Images of the 1.3~mm continuum emission for Cepheus~A in each of the four Stokes 
parameters taken with CARMA. The Stokes $I$ emission appears in the four panels in grayscale.
Overlayed yellow contours for the Stokes $I$ are at -3, 3, 6, 9, 15, 25, 35 $\times\sigma$, 
where $\sigma=10$\mjy is the rms noise level of the image. For Stokes $Q$, $U$ and $V$, red contours 
are at -3, 3, 4, 5, 6 $\times\sigma$, where $\sigma=1$\mjy is the rms noise level of these 
three images. The two main sources in the field are labeled in the Stokes $I$ image. The 
orientation of the disk-like structure ($P.A.\sim135\degr$) and the jet of HW2 ($P.A.\sim45\degr$) are marked with blue lines (dashed and solid, respectively). The synthesized beam is shown in the bottom right corner.}
\label{fall}
\end{figure*}

Figure \ref{fall} shows a set of images of Cepheus A HW2 in full Stokes 
1.3~mm continuum emission. Two bright sources are detected in Stokes $I$, 
known as HW2 and HW3c in the literature. One more feature is detected at the 
3$\sigma$ level, which is located $1\farcs5$ southeast of HW2; this feature is 
probably due to a peak in the noise spectrum. As reported in previous 
submillimeter observations \citep{2005Patel,2007Torrelles}, HW2 has two 
protrusions northwest and southeast of the protostellar location: SMA1 and 
SMA3. These features have been interpreted as possible 
protostellar cores in a tight cluster, but also have been suggested to be part of a 
more extended structure associated with the disk \citep{2007Brogan}. 
We measured a total integrated flux of 893~mJy for HW2. 
A 2D-Gaussian fit to its emission (Table \ref{tcont}) gives a structure 
with a deconvolved size $1\farcs21\times0\farcs82$ ($850\times575$~AU at 700~pc).
The P.A. of the HW2 millimeter emission is $135\degr$, which is orthogonally oriented 
to the thermal jet \citep[P.A.$=44\degr\pm4\degr$, ][]{1996Torrelles}, thought to 
be the current ejection of the episodic precessing outflow \citep{2009Cunningham,2013Zapata}. 
The deconvolved size obtained from this fit is slightly larger than reported by 
\cite{2005Patel} from submillimeter observations, maybe due to a better 
sensitivity of CARMA to extended emission or differences in the spatial distribution 
of grains of different sizes. The peak location of the HW2 continuum 
source agrees with the protostar position derived in \cite{2006Curiel}. HW3c is located 
$3\farcs5$ south of HW2. Its integrated 1.3~mm flux is 112~mJy,
and a 2D-Gaussian fit shows that it is extended, with a deconvolved size of $1\farcs4\times0\farcs7$ 
($980\times490$~AU), and elongated in the north-south direction 
(P.A.$=2\degr\pm10\degr$). Its peak position agrees with the position of the 1.3~cm 
radio source reported by \cite{1996Torrelles}. 

\begin{figure*}[!h]
\epsscale{1}
\plotone{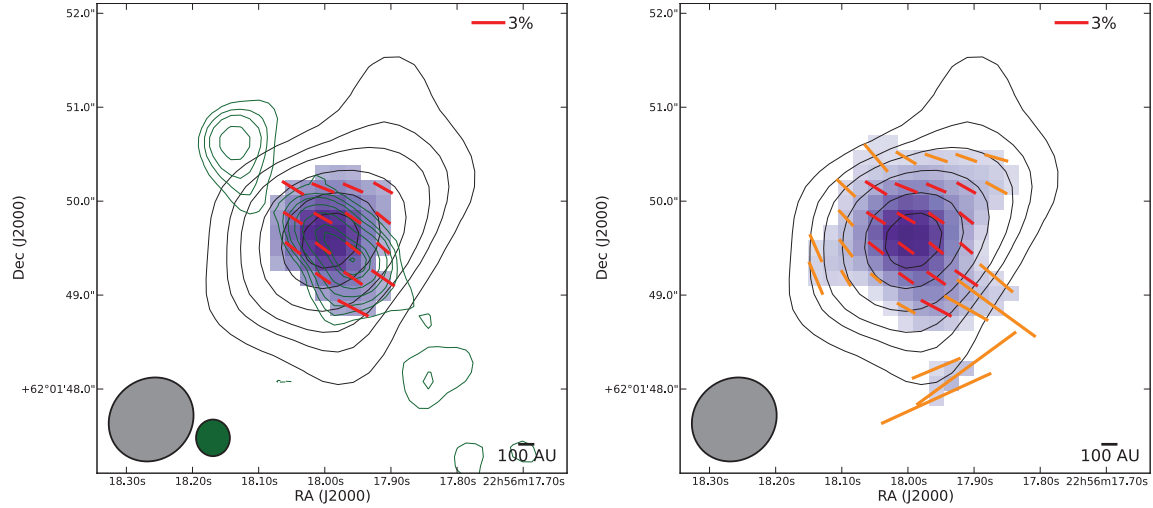}
\caption{\textbf{Left:} Polarimetric map (polarization E-vectors) of the Cepheus~A~HW2 disk-like structure. 
Fractional polarization red vectors over $3\sigma$ displayed. Black contours are Stokes $I$ data with the 
same levels as in Figure \ref{fall}. Green contours correspond to the VLA C~band continuum emission 
showing the radio jet at -6, -3, 3, 6, 10, 20, 40, 80, 100 and 140 times the rms noise level of the 6~cm 
radio map, $28~\mu$Jy~beam$^{-1}$) associated with the HW2 massive protostar.
Color scale shows the polarized intensity over $3\sigma$. The 
synthesized beam of both CARMA (grey) and VLA (green) 
observations are shown at the bottom left corner. \textbf{Right:} Same map as in left panel but 
including orange vectors between $2\sigma$ and $3\sigma$. The polarized intensity (color scale) 
also shows emission over $2\sigma$.}
\label{fpoli}
\end{figure*}

The images of the Stokes $Q$ and $V$ parameters are devoid of significant emission 
(Figure \ref{fall}). However, the Stokes $U$ image shows a clear detection ($>7\sigma$) 
peaking at the HW2 location. Hence, the linear polarized intensity image (Figure 
\ref{fpoli}) resembles the Stokes U emission distribution, and the polarized angle 
dispersion throughout HW2 is small. No other source was found to be polarized 
at a $3\sigma$ level (3\mjy) within the field of view. The polarization 
fraction for HW3c is less than 1.9\%. For HW2, we measured a total linear 
polarized intensity of $9\pm1$~mJy. The CARMA observations partially resolved 
the HW2 polarized emission, which spreads over two independent beams (5-6 beams at the 
$2\sigma$ level), with an average polarization fraction of $2.0\%\pm0.4\%$. 
The morphology of the polarized emission shows two preferential elongations: 
one roughly aligned with the HW2 radio jet (northeast-southwest of the 
protostar location) and the other aligned with the major axis of the disk-like structure 
and due northwest. 
The average orientation of the electric field vectors ($57\degr\pm6\degr$, measured 
counterclockwise from north) is deviated about $12\degr$ with respect to the HW2 radio 
jet and the disk normal directions (as shown in the left panel of Figure \ref{fpoli}), but the deviation is smaller 
when considering the orientation of the northern 
lobe of the radio jet alone \citep[P.A.$=50\degr$, ][]{2006Curiel}. Moreover, the 
polarization vectors (P.A.$=57\degr\pm6\degr$) are not exactly perpendicular 
to the dusty disk-like structure orientation (P.A.$=135\degr$).

\section{Modeling and Discussion}
In this section we discuss the nature of the polarimetric observations resolving 
the disk of Cepheus~A~HW2. First, we summarize the interpretations given to similar
observations in other protostellar disks. It is of utmost importance to figure out the dominant
polarization mechanism to analyze these data since each mechanism provides different 
constraints on the disk's dust properties. We fit three polarization models to the data
and comment on the results here. 

\subsection{Previous polarized observations resolving disks of low-mass protostars}
In the disk of HL~Tau, the polarization vectors are uniformly oriented and parallel to 
the minor axis, but the polarized emission stems mainly from an elongated region along 
the disk major axis \citep{2014Stephens}. Simple models for direct emission do not 
completely agree with the observations. This disagreement motivated new theoretical studies 
that showed  that dust scattering is a viable alternative to explain the nature of the millimeter 
polarized emission of the HL~Tau disk \citep{2015Kataoka,2016Yang1,2016Kataoka}.
For the disk of L1527, the 1.3~mm 
CARMA polarization vectors are again uniformly oriented parallel to the minor axis of the disk, but 
the polarized emission seems to be along the same axis \citep{2015SeguraCox}. These 
characteristics are expected for the direct emission of magnetically aligned grains threaded 
by a toroidal field in an edge-on disk \citep{2016Yang2}. However, the higher than average 
polarization fraction may also agree with the expectation for scattering in high-inclination disks.
In the NGC1333~IRAS4A1 disk, the JVLA\footnote{Jansky Very Large Array} polarized 
8~mm/1~cm dust continuum emission \citep{2015Cox} has been interpreted as possibly 
due to a mixture of magnetically aligned grains embedded in a toroidal magnetic field 
in the disk and dust scattering \citep{2016Yang2}. For the Class~0 disk IRAS~16293-2422~B
\citep[SMA\footnote{Submillimeter Array} observations, ][]{2014Rao}, the geometry 
of the disk (face-on) and the complex morphology of the $870~\mu$m polarization vectors 
(swirled and curved) could be explained by a toroidal magnetic field model.

\subsection{Polarization vector models}
In order to better understand the nature of the polarized emission found in the disk-like structure 
of Cepheus~A~HW2, we test the orientation of the polarization vectors 
with three different models: grains aligned with a uniform magnetic field, grains aligned
with a toroidal magnetic field, and polarization due to scattering. We fit each model to the 
observed orientations of the polarization vectors, but we did not fit the spatial distribution of the polarized emission 
nor the polarization fraction.

\subsubsection{Uniform magnetic field model}
The first model that we test is one with grains aligned with a uniform magnetic field threading 
the disk-like structure approximately 
along its major axis, which would cause a distribution of polarization vectors 
perpendicular to the field (i.e., northeast-southwest, along the minor axis). This distribution 
of vectors has the same morphology as the large-scale $850~\mu$m polarization observations 
\citep{2009Matthews}, which traces a northeast-southwest parsec scale 
hour-glass shaped magnetic field, with a uniform distribution threading the central protostar(s),
including HW2. In general, field lines are expected to be toroidal wrapped due to 
the rotation of the gas in accretion disks, but it is feasible that in early stages and far from 
the inner disk, the magnetic field can be almost uniform.

\begin{figure*}[!h]
\epsscale{1}
\plotone{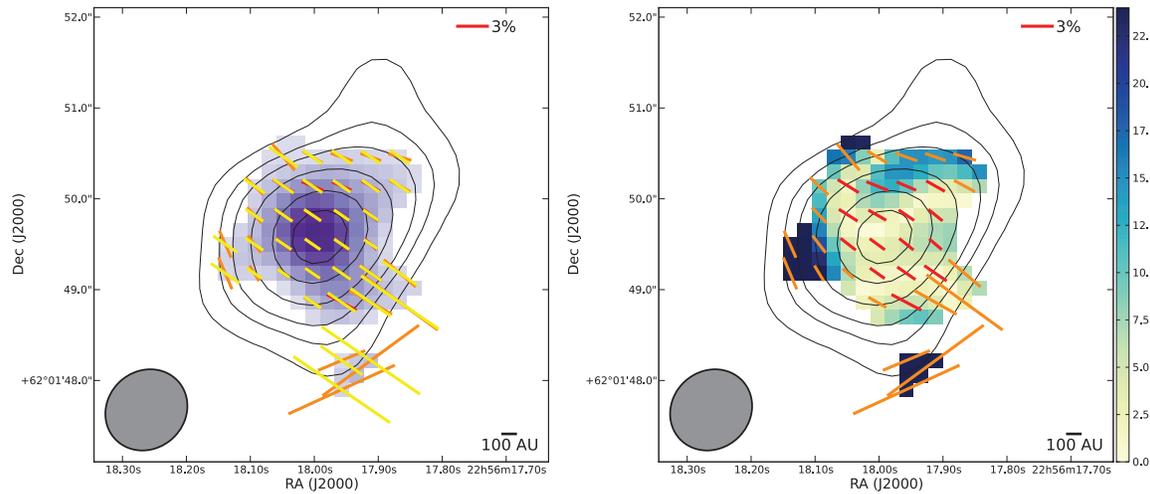}
\caption{\textbf{Left:} Uniform magnetic field model (yellow polarization E-vectors) compared with the 
1.3~mm polarimetric observations (orange and red vectors are CARMA detections between 
$2\sigma$ and $3\sigma$ and $>3\sigma$, respectively). Contours are Stokes $I$ data with the 
same levels as in Figure \ref{fall}. Color scale shows the polarized intensity over $2\sigma$. 
\textbf{Right:} The absolute value of the residuals between the observed and the modeled orientations of the polarization 
vectors (color scale, in degrees). The mean rms level of the E-vectors is $6\degr$. Contours are 
Stokes $I$ data with the same levels as in Figure \ref{fall}. Vectors are as in the right panel of Figure \ref{fpoli}.
}
\label{funif}
\end{figure*}

We make a weighted linearly fit to the 1.3~mm polarization observations (orientations of the polarization vectors only) with a uniform field (Figure \ref{funif}). The average fit orientation of the uniform field is given by a polarization 
vector set with P.A.$=56\degr$, implying a magnetic field oriented with a P.A.$=146\degr$. 
This is roughly consistent with the large-scale P.A. of $135\degr$. As indicated by the 
residuals (mostly below the $3\sigma$ threshold of $18\degr$), the model is a good 
match to the observed data.

\subsubsection{Toroidal magnetic field model}
The second model that we test is a simple one, with grains aligned with a toroidal magnetic 
field. Toroidal fields are proposed for MRI (magneto-rotational instability) dominated 
disks of low-mass protostars \citep[e.g.,][]{2007Cho}. It is therefore an appropriate 
model to test against the CARMA observations. 

\begin{figure*}[!h]
\epsscale{1}
\plotone{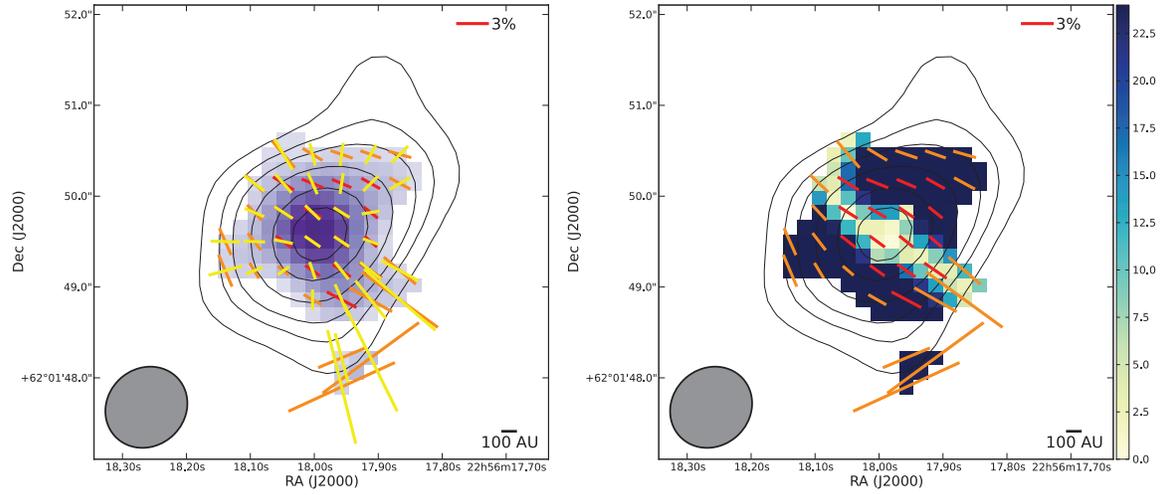}
\caption{Same as Figure \ref{funif}, except for the toroidal magnetic field model (see text).}
\label{ftoroid}
\end{figure*}

We fit the observations (orientation of the polarization vectors only) allowing two free parameters: inclination 
($i$) with respect to the plane of the sky ($0\degr$ means face-on and $90\degr$ is edge-on) 
and position angle (more details of the model are given in the Appendix). The best fit gives 
$i=45\degr$ and P.A.$=141\degr$ (Figure \ref{ftoroid}). 
A visual comparison with the uniform model fit clearly shows that the toroidal field is not a
perfect match to the observations, with most of the residuals over $3\sigma$, except for the 
region around the minor axis of the disk-like structure.

\subsubsection{Scattering in an inclined disk}
\begin{figure*}[!h]
\epsscale{1}
\plotone{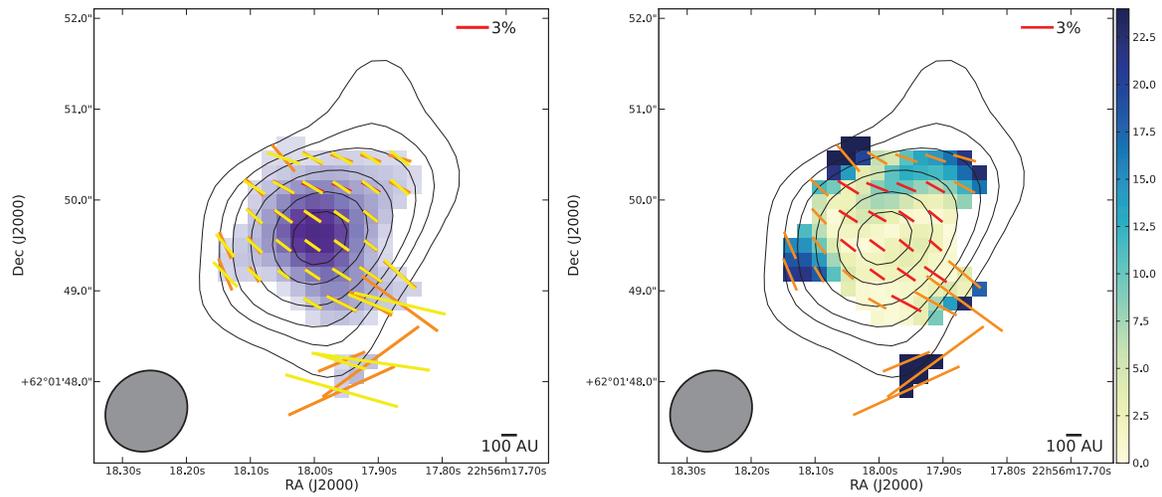}
\caption{Same as Figure \ref{funif}, except for the scattering model (see text).}
\label{fscatter}
\end{figure*}

For the third model, we test inclination-induced scattering as the polarization mechanism 
\citep{2016Yang1}. We used a phenomenological model which is described in the Appendix. 
This is strictly a morphological comparison of the orientations of the polarization vectors for the model and 
data, and we neglect comparing the polarized intensity morphology. We neglect the latter
to reduce the free parameters, given the limited amount of independent beams detected 
across the source. The best fit is found when the inclination is 
$i=61\degr$ and the P.A.$=145\degr$ (left panel of Figure \ref{fscatter}). 
\cite{2005Patel} estimated an inclination of $62\degr$ and a P.A. of $135\degr$ for 
the gaseous and dusty disk, which agrees with these results. The residuals of this 
model (right panel of Figure \ref{fscatter}) are in general below the $3\sigma$ threshold, 
even lower (mean of the absolute residuals is $7\degr$) than those obtained for the
uniform magnetic field model (mean of the absolute residuals is $8\degr$).

\subsection{Morphology of the polarized intensity}
Our models do not include the spatial distribution of the polarized intensity emission. If 
the morphology of the polarized emission is elongated along the major or minor 
axis of a disk, this can indicate
that scattering or direct emission dominates the polarized emission, 
respectively \citep[][]{2016Yang2}. A combination of both mechanisms produces also
elongated polarized intensity along the major axis. 
In the case of HW2, the polarized intensity seems to be slightly elongated in the 
northeast-southwest direction (deconvolved size of 
$1\farcs2\pm0\farcs2\times1\farcs0\pm0\farcs2$ with a P.A. between $20\degr$ 
and $30\degr$, depending on the vectors between $2\sigma$ 
and $3\sigma$ are included or not), i.e., elongated across the minor axis of the disk-like structure.
Since the elongation of the polarized intensity cannot be accurately determined from the present observations, we do not derive any convincing conclusion about the polarization mechanism only from the intensity distribution.

\subsection{Interpretation of the data}
The modeling of the orientations of the polarization vectors indicates a good match for the uniform 
magnetic field and scattering scenarios. It is clear from the residuals of Figures 
\ref{funif}-\ref{fscatter} that the toroidal field model is significantly worse than the 
other two models. Nevertheless, the 
polarized intensity shows a slight elongation along the minor axis of the disk-like structure, 
which could favor the direct emission mechanism from a toroidal field over the scattering. 

We must make note of the caveats involved with this method of determining 
possible polarization mechanisms. First, CARMA is resolving the polarized emission only on 
a limited number of independent beams across the source.
Both scattering and direct emission from grains aligned by a uniform magnetic field
produce a similar pattern of vectors at the center of the disk-like structure, which explains the
observations well, with residuals below the $3\sigma$ level. The lack of higher angular 
resolution and sensitivity also hinders the analysis 
of the elongation of the polarized intensity. When considering the emission over $3\sigma$ 
(Figure \ref{fpoli} left), it seems slightly elongated along the minor axis, but when considering 
the emission over $2\sigma$ (Figure \ref{fpoli} right), this preferential orientation is less clear. 
It is therefore complicated to use the elongation of polarization to distinguish between the 
direct emission and the scattering models.

Second, we are probably dealing with a massive disk-like structure that may be optically 
thick, particularly at the center. Note that our models are based on a 2D flatten disk and
should be review to appropriately address how the optical depth modifies the polarization
due to direct emission and/or scattering.

Finally, it is unclear that the millimeter emission toward HW2 comes
entirely from a circumstellar accretion disk. The disk-like 
structure of HW2 is considered a dust and molecular rotating disk 
\citep{2005Patel}, but the dust emission
could also contain a common envelope for the protostars, irregularities due to substructure 
of a multiple system, or even morphological disturbances produced by the action of the
precessing jet. When present observations are improved, the assumed scattering model 
should probably be refined to include these effects. 

The CARMA observations presented here cannot alone distinguish between the uniform magnetic field
(direct emission) and the scattering scenarios, assuming the accretion disk dominates the dust emission 
in HW2.
On one hand, the 
uniform magnetic field would imply the presence of a large-scale field possibly dragged by 
the dust and gas accretion into a flatten structure which has yet to wrap into a toroidal field, at least 
at large scales far away from the protostar(s). On the other hand, the scattering hypothesis
would imply that grains have grown up to $\sim10$s of microns \citep{2016Yang1,2016Kataoka}. 
Grains grow very rapidly from $\sim1~\mu$m (ISM grain sizes) to several 
$10$s of microns \citep[grains in Class~0 and Class I disks,][]{2009Kwon,2012Chiang,2014Miotello}. 
Since the age of HW2 has been estimated  to be 
$10^4$~yr \citep[from its outflow dynamical time,][]{2009Cunningham}, this could be a new 
evidence of fast grain growth in dense environments. In the end, it may be a combination 
of polarization due to both mechanisms in play, although the hybrid (with a simple magnetic field configuration) 
model proposed by \cite{2016Yang2} may not perfectly explain the observations either.

\section{Summary and Conclusions}
We have reported the first resolved polarization observations of the disk-like structure of the 
high-mass protostar Cepheus~A~HW2. The 1.3~mm CARMA data show a partially resolved 
distribution of polarization vectors, suggesting an uniform morphology toward the center 
of the disk-like structure. 
The two likely mechanisms which could lead to the polarization of the millimeter
emission are: (1) a magnetic field aligning the grains that emit polarized radiation, or (2) 
scattering from large enough dust grains (10s of microns). Fitting the observations with two 
phenomenological models (the uniform magnetic field threading the disk-like structure and the 
scattering in the inclined disk), we find both of these scenarios to be plausible and indeed, 
taking into account factors such as the opacity, a mixture of them may be expected. 

If the main cause of the polarization is the grain alignment via magnetic forces, future intermediate
resolution SMA observations could fill the gap between the CARMA ($1\arcsec$ resolution) and the
previous JCMT ($20\arcsec$ resolution) millimeter and submillimeter data. This is crucial to make 
the link between the large scale hour-glass shape morphology and the uniform vectors threading the 
most dense part of the disk-like structure. It is also clear from our fitting that the toroidal magnetic field does
not match the CARMA observations. On the contrary, if the dominant cause of the polarized emission 
is scattering, the grains in the disk should have rapidly grown to 
sizes at least of the order of $\sim100~\mu$m 
\citep[probably in a few  $10^4$~yr,][]{2015Kataoka,2016Yang1}. This would add new 
evidence for quick grain growth taking place in dense disk-like structures, even in the adverse 
environment of a small cluster of massive protostars, where radiation or dynamical effects could 
restrict the growth of planetesimals. Both, multi-wavelength and sub-arcsecond angular resolution 
observations would give details to better distinguish which is the dominant 
mechanism.

\acknowledgments
We thank all members of the CARMA staff that made these observations possible. 
Support for CARMA construction was derived from the Gordon and Betty Moore Foundation, the Kenneth T. and Eileen L. Norris Foundation, the James S. McDonnell Foundation, the Associates of the California Institute of Technology, the University of Chicago, the states of Illinois, California, and Maryland, and the National Science Foundation. Ongoing CARMA development and operations are supported by the National Science Foundation under a cooperative agreement, and by the CARMA partner universities. J.M.G. acknowledges support from MICINN AYA2014-57369-C3-P, the MECD PRX15/00435 (Spain), and SICGPS \lq\lq Magnetic Fields and Massive Star Formation\rq\rq (USA) grants. L.~L. is supported in part by the NSF grant 1139950. S.C. acknowledges support from DGAPA, UNAM, and CONACyT, M\'exico. E.C. and S.P.L. are thankful for the support of the Ministry of Science and Technology(MoST) of Taiwan through Grant MoST 102-2119-M-007-004-MY3. 

Facilities: \facility{CARMA}
\bibliography{biblio}

\appendix
In this section, we describe the expressions used to generate the synthetic polarization
electrical vector images of the toroidal and scattering models 
(see Modeling and Discussion section). To make an E-vector image we first create
its corresponding Stokes $Q$ and $U$ images (Miriad tasks IMGEN and MATHS) 
and convolve them with the CARMA synthesized beam (Miriad task CONVOL). We produce 
the E-vector image using Miriad task IMPOL. This image is then rotated by a position 
angle (task MATHS), since the Stokes $Q$ and $U$ images are originally obtained 
for P.A.$=0\degr$. 

For the toroidal field model (radial E-vectors), we implement the following expressions
for a 2D Cartesian coordinate system $(x,y)$ in the plane of the sky, with the  
x- and y-axes along the major and minor axes of the disk-like structure 
respectively, and with the origin at the disk center:

\begin{equation}
Q=-D\left(x,y,i_2\right)\cdot \frac{x^2-\left(y\cos{i_2}\right)^2}{x^2+\left(y\cos{i_2}\right)^2}\quad,
\label{eqtq}
\end{equation}

\begin{equation}
U=-D\left(x,y,i_2\right)\cdot \frac{2xy\cos{i_2}}{x^2+\left(y\cos{i_2}\right)^2}\quad,
\label{eqtu}
\end{equation}
  
where 
\begin{equation}
D\left(x,y,i_1\right)=F_0\cdot\left(\frac{\sqrt{x^2+\left(y\cos{i_1}\right)^2}}{r_0}\right)^{-\alpha}\cdot e^{-\left(\sqrt{x^2+\left(y\cos{i_1}\right)^2}/r_0\right)^{\beta}}\quad.
\label{eqD}
\end{equation}

In the previous equations, $D\left(x,y,i\right)$ is a 2D function 
that is fitted to the Stokes $I$ image (avoiding the $x=0$, $y=0$ point, which we set as the position 
observed Stokes $I$ peak). In the fitting process, the free parameters are the 
exponents $\alpha$ and $\beta$, a characteristic radius $r_0$, the parameter $F_0$ (related to
the central flux density of the disk-like structure) and the inclinations with respect to the plane of the sky $i_1$ and $i_2$. 
P.A.$_1$ and P.A.$_2$ (the angle of the disk in the plane of the sky counterclockwise from North) 
are also free parameters, but they do not appear explicitly 
in the above expressions, since they are implemented as a later rotation of the E-vector image. 
$D\left(x,y,i\right)$ is reminiscent of the standard viscous accretion disk 
model \citep{1981Pringle} used by other authors to fit the continuum emission of an accretion disk 
\citep[e.g.,][]{2011Kwon, 2016Yang1}. The fit of Equation \ref{eqD} to the Stokes $I$ data is a 
sufficient representation of the disk-like structure emission, with residuals below the $3\sigma$ 
level. The fit values for the different parameters are presented in Table \ref{tdisk}. After fitting the
disk continuum emission the obtained values of $F_0$, $r_0$, $\alpha$ and $\beta$ are fixed and
used to create the Stokes $Q$ and $U$. The obtained values of $i_1$ and P.A.$_1$ are discarded and
left as two new free parameters ($i_2$ and P.A.$_2$) in Equations \ref{eqtq}, \ref{eqtu}, \ref{eqsq} and \ref{eqsu}. 
By solving for $i_2$ and P.A.$_2$, we can find the best possible fit for each polarization model. This gives us an additional qualitative check on the models, since we can compare the fitted disk inclination and position angle from the orientations of the polarization vectors ($i_2$, P.A.$_2$) to that reported in the literature and the obtained from the fit of the Stokes I emission ($i_1$, P.A.$_1$).

The following mathematical expressions for the scattering model where inspired by equations (16) and (17) 
of \cite{2016Yang1}. Our model is a phenomenological function and hence it is not based
on physical principles. However, these expressions provide a representation of  the effects of disk 
inclination on the orientations of the polarization vectors of scattered radiation shown in Figure 2 of \cite{2016Yang1}. 
\begin{equation}
Q=D\left(x,y,i_2\right)\cdot \left(1-\sin^2{i_2}\cdot\frac{x^2}{x^2+y^2}\right)\cdot
\frac{x^2\cos^2{i_2}-y^2}{x^2\cos^2{i_2}+y^2}\quad,
\label{eqsq}
\end{equation}

\begin{equation}
U=D\left(x,y,i_2\right)\cdot \left(1-\sin^2{i_2}\cdot\frac{x^2}{x^2+y^2}\right)\cdot
\frac{2xy\cos^2{i_2}}{x^2\cos^2{i_2}+y^2}\quad.
\label{eqsu}
\end{equation}

\begin{deluxetable*}{cccccccc}
\tablewidth{0pt}
\tablecolumns{8}
\tablecaption{Parameters of the Serpens South 3~mm continuum sources}
\phs
\tablehead{
\colhead{Source} & \colhead{Right Ascension} & \colhead{Declination} & \colhead{Stokes I Peak} & \colhead{Stokes I Flux} & \colhead{P\tablenotemark{(a)}} & P\%\tablenotemark{(b)} & \colhead{P.A.\tablenotemark{( c)}} \\
\colhead{}  & \colhead{J2000} & \colhead{J2000} & \colhead{(mJy beam$^{-1}$)}  & \colhead{(mJy)} & \colhead{(mJy)} & \colhead{(\%)} & \colhead{}
}
\startdata
HW2  & $22^h56^m17\fs989$ & $62\degr01\arcmin49\farcs60$ & 423$\pm$10 & 863$\pm$20 & 9$\pm$1  & 2.0$\pm$0.4 & 57$\pm$6\\
HW3c & $22^h56^m17\fs976$ & $62\degr01\arcmin46\farcs10$ & 68$\pm$10 & 160$\pm$6 & \nodata & \nodata & \nodata \\

\enddata 
\tablecomments{}
\tablenotetext{(a)}{Polarized intensity flux.}
\tablenotetext{(a)}{Fraction of polarized emission (percentage average).}
\tablenotetext{(c)}{Position angles of the polarized emission are measured counterclockwise.}

\label{tcont}
\end{deluxetable*}

\begin{deluxetable}{cc}
\tablecolumns{2}
\tablecaption{Parameters obtained from the Stokes~$I$ fit}
\phs
\tablehead{
\colhead{Parameter} & \colhead{Fit} 
}
\startdata
$F_0$\tablenotemark{(a)} & $2.5\times10^{-4}$~Jy~pixel$^{-1}$ \\
$r_0$ & $1\farcs0$ \\
$\alpha$  & 1.2 \\
$\beta$ & 1.6 \\
$i_1$ & $51\degr$ \\
P.A.$_1$ & $142\degr$ \\
\enddata 
\tablenotetext{(a)}{The size of a pixel is $0\farcs05$.}
\label{tdisk}
\end{deluxetable}


\end{document}